\documentclass[apj]{emulateapj}
\usepackage{graphicx}
\usepackage{amssymb}
\usepackage{epstopdf}
\usepackage{amsmath}
\usepackage{natbib}
\usepackage{subfigure}
\usepackage{mathtools}
\usepackage{paralist}
\usepackage{hyperref}

\def\beq{\begin{equation}}
\def\eeq{\end{equation}}
\def\beqa{\begin{eqnarray}}
\def\eeqa{\end{eqnarray}}
\def\bseq{\begin{subequations}\begin{eqnarray}}
\def\eseq{\end{eqnarray}\end{subequations}}

\def\kperp{k_{\bot}}
\def\kpar{k_{\|}}
\def\kt{k_{\tau}}
\def\k{{\bf k}}

\def\th{{\mathbf{\theta}}}

\def\u{{\bf{u}}}

\def\I{{I}}

\def\Ih{{\hat{\I}}}

\def\r{{\bf{r}}}

\def\r{{\bf{r}}}

\def\Pkt{{\hat{P}(\kperp,\kt)}}
\def\Pkp{{\hat{P}(\kperp,\kpar)}}
\def\Ph{{\hat{P}}}

\def\eppsilon{{$\varepsilon$ppsilon}}

\begin{document}
\title{Understanding the Diversity of 21~cm Cosmology Analyses}
\author{Miguel F. Morales\altaffilmark{1,2}}
\author{Adam Beardsley\altaffilmark{3}}
\author{Jonathan Pober\altaffilmark{4}}
\author{Nichole Barry\altaffilmark{5,6}}
\author{Bryna Hazelton\altaffilmark{1,7}}
\author{Daniel Jacobs\altaffilmark{3}}
\author{Ian Sullivan\altaffilmark{8}}
\altaffiltext{1}{Department of Physics, University of Washington, Seattle, 98195}
\altaffiltext{2}{Dark Universe Science Center, University of Washington, Seattle, 98195}
\altaffiltext{3}{School of Earth and Space Exploration, University of Arizona, 85287}
\altaffiltext{4}{Department of Physics, Brown University, Providence, 02912}
\altaffiltext{5}{School of Physics, University of Melbourne, Victoria, 3010}
\altaffiltext{6}{ARC Centre of Excellence for All Sky Astrophysics in 3 Dimensions (ASTRO 3D)}
\altaffiltext{7}{eScience Institute, University of Washington, Seattle, 98195}
\altaffiltext{8}{Department of Astronomy, University of Washington, Seattle, 98195}


\begin{abstract}

21~cm power spectrum observations have the potential to revolutionize our understanding of the Epoch of Reionization and Dark Energy, but require extraordinarily precise data analysis methods to separate the cosmological signal from the astrophysical and instrumental contaminants. This analysis challenge has led to a diversity of proposed analyses, including delay spectra, imaging power spectra, m\nobreakdash-mode analysis,  and numerous others. This diversity of approach is a strength, but has also led to confusion within the community about whether insights gleaned by one group are applicable to teams working in different analysis frameworks. In this paper we show that all existing analysis proposals can be classified into two distinct families based on whether they estimate the power spectrum of the {\it measured} or {\it reconstructed} sky. This subtle difference in the statistical question  posed largely determines the susceptibility of the analyses to foreground emission and calibration errors, and ultimately the science different analyses can pursue. In this paper we detail the origin of the two analysis families, categorize the analyses being actively developed, and explore their relative sensitivities to foreground contamination and calibration errors.

\end{abstract}

\maketitle

\section{Introduction}

Radio observations of redshifted 21~cm hydrogen emission have the potential to reveal our Cosmic Dawn, constrain Dark Energy, and constrain cosmological parameters over an unprecedented cosmic volume. Major power spectrum (PS) analyses of hundreds of hours of data are underway by PAPER, LOFAR, GMRT, GBT, MWA and CHIME, and HERA is about to start science data collection (Donald C. Backer Precision Array for Probing the Epoch of Reionization, \citealt{Parsons2010}; Low-Frequency Array, \citealt{Haarlem2013}; Giant Metrewave Radio Telescope, \citealt{Gupta2017}; Green Bank Telescope, \citealt{Prestage2009}; Canadian Hydrogen Intensity Mapping Experiment, \citealt{Bandura2014}; Murchison Widefield Array, \citealt{Tingay2013}; Hydrogen Epoch of Reionization Array, \citealt{DeBoer2016}). Many different power spectrum analyses have been proposed, and a number have been developed into full fledged data analysis pipelines.

The community is learning rapidly, and instrument/analysis features such as the `foreground wedge,' the `window,' and the `pitchfork' are major focuses of the literature (see \S\ref{sec:chromaticity} and references therein). However, as the effects are always measured or simulated within one of the many analysis frameworks, the details of the analysis become intertwined with the results. Consequently, even domain experts are often uncertain which effects identified by other groups will apply to their own work.

After a careful review of the 21~cm cosmology literature, we find that all of the proposed analyses can be sorted into two `families' based on the mathematics of the final estimator (\S\ref{sec:estimators}). The first family of analyses calculates the PS of the {\it measured} sky, while the second family calculates the PS of the {\it reconstructed} sky. This subtle difference in the statistical question being posed has profound effects on the sensitivity of the analyses to foreground emission and calibration errors.

The mathematics for how foregrounds contaminate measured (delay-style) and reconstructed (imaging-style) power spectra is fully developed in a paper by \cite{Liu2014a}. However, the practical ramifications of this distinction have not been fully explored and are under-appreciated by the community. The goal of this paper is to clearly illustrate the connections between different analyses. This will both enable practitioners to recognize which effects apply to their efforts, and illuminate the pros and cons of measured and reconstructed power spectra. 

After a brief review of interferometric 21~cm measurements with chromatic instruments (\S\ref{sec:chromaticity}), we identify the two families of analysis (\S\ref{sec:estimators}). We then use this insight to explore how astrophysical foregrounds (\S\ref{sec:foreground}) and calibration errors (\S\ref{sec:calibration}) appear in the two families of analyses. In \S\ref{sec:ProsCons} we discuss the pros and cons of the two approaches, and where we see the state-of-the-art 21~cm PS analysis heading in the years to come.

\section{Instrument chromaticity}
\label{sec:chromaticity}

All interferometric measurements are inherently chromatic. Because the astrophysical foregrounds are very spectrally smooth, in principle the line-of-sight 21~cm fluctuations are free of foreground contamination (see \citealt{Furlanetto:2006bq} and \citealt{Morales:2010cb} for reviews). However, the natural chromatic response of an instrument leads to a wedge-like region of contamination (predicted in \citealt{Datta:2010he,Morales:2012uk,Vedantham:2011wu,Parsons2012,Trott:2012fb,Hazelton:2013fu,Thyagarajan:2013cs,Dillon2013,Thyagarajan2015}  and observed in \citealt{Pober:2013cp,Dillon:2013uc,Thyagarajan2015a,Jacobs2016,Beardsley2016}). The origin of this contamination can be seen by carefully examining the frequency response of a single visibility. In this section we briefly review the origin of chromaticity and the associated wedge and window in 21~cm cosmology PS, closely following the developments in \citet{whitebook,TMS,Morales2012,Liu2014,Pober2016}.

The measured visibilities are the cross-correlations of the electric field measured between pairs of antennas. The visibility for one pair of antennas can be written as 
\beqa
%
v_{ij}(f) =  \int \Big\langle A_i(\th,f)E(\th,f) &  \nonumber \\ 
\times A^*_j(\th,f)E^*(\th,f) & e^{-i2\pi [\Delta \r_{ij}\, f/c] \cdot \th}\Big\rangle_{\!t} \ d^2\th, 
\hbox{}
\label{eq:vimage}
\eeqa
where $E(\th,f)$ is the instantaneous electric field of the sky as a function of direction and frequency, $A$ is the directional electrical sensitivity of antennas $i$~\&~$j$, and $\Delta \r_{ij}$ is the physical distance between the antennas. The observed electric field detected by each antenna is given by $AE$ with the exponential describing the direction-dependent time delay of the signal at the second antenna relative to the first (the Green's function propagator of the electric field for a far field source). The measured visibility is then the angular integral and time average (angle brackets) of the cross-power between the two antennas. 

We can move from electric field to brightness description by using $\langle E E^* \rangle_t = I$. We can further simplify the resulting equation by assuming the antennas have the same angular response so we can use the antenna power sensitivity $B = |A|^2$, and by replacing the physical distance with the antenna separation in wavelengths $\u_{ij} = \Delta \r_{ij}\, f/ c$:
%
\beq
v_{ij}(f) = \int I(\th,f)B(\th,f)e^{-i2\pi \u_{ij}(f) \cdot \th} d^2\th.
\label{eq:vimage2}
\eeq 
Note both $\u$ and $\theta$ are vectors in this notation. We can immediately see that for a single flat spectrum source ($I$ is a $\delta$-function at $\th_S$) the visibility will be proportional to 
$B(\th,f)e^{-i2\pi \u_{ij}(f) \cdot \th_S}$---the visibility will oscillate with increasing frequency. The longer the baseline length $\u_{ij}$, the more rapid the fequency oscillations. While there are several sources of instrument chromaticity, fundamentally the array appears larger (in wavelengths) at higher frequencies.

To understand the effect of this visibility oscillation on the power spectra, it is useful to continue to the Fourier representation of the instrument measurement. The integral in Equation~\ref{eq:vimage2} is equivalent to a 2D angular Fourier transform from $\th \Rightarrow \u$. Taking the Fourier transform we obtain
\beq
v_{ij}(f) = \int \big[ I(\u,f)*B(\u,f) \big] \delta(\u - \u_{ij}')\, d^2\u,
\label{eq:vu1}
\eeq
where we have Fourier transformed both the sky brightness $I$ and the antenna brightness sensitivity $B$,\footnote{Often Fourier Transformed variables are indicated with a tilde (e.g.\ $\tilde{I}(\u,f)$), but as we want to use a hat to differentiate between true and estimated quantities later in the paper we omit the tilde and indicate the Fourier Transform by the variable being a function of the Fourier coordinate $\u$.} used the Fourier convolution theorem ($*$ is the convolution operator), and expressed the discreet baselines $\u_{ij}$ as a $\delta$-function. Alternatively we can use the $\delta$-function to remove the convolution over all $\u$ to obtain
\beq
v_{ij}(f) = \int I(\u,f)B(\u - \u'_{ij},f)\, d^2\u.
\label{eq:vu2}
\eeq
In words, the visibility is equal to the Fourier Transform of the sky brightness integrated by the compact uv-antenna response pattern $B(\u)$. 

\begin{figure}
\begin{center}
\includegraphics[width=\columnwidth]{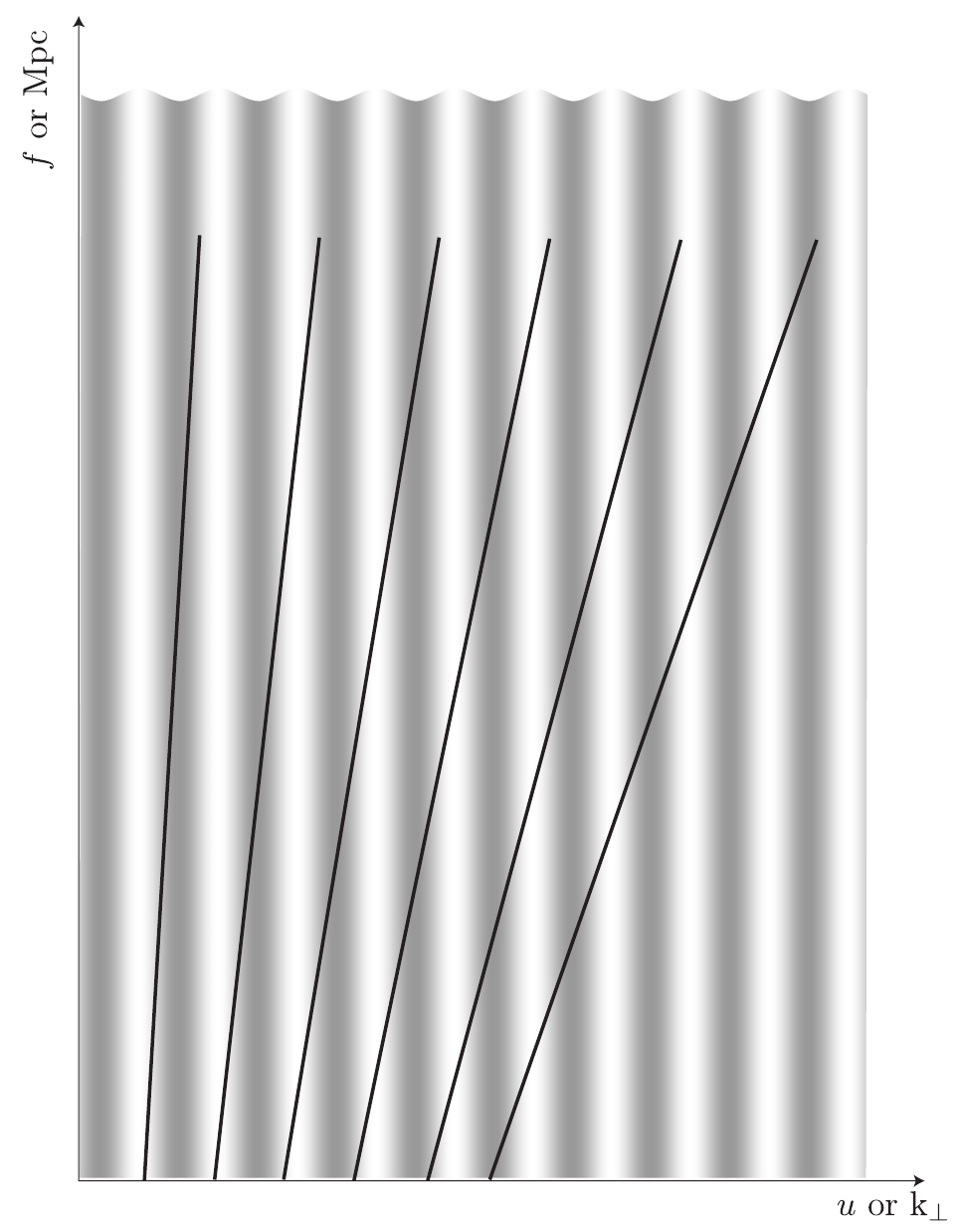}
\caption{Cartoon adapted from \citet{Morales:2012uk} showing how flat spectrum foregrounds create oscillatory visibilities, and why the oscillations are faster for longer baselines. Described in detail in \S\ref{sec:chromaticity}.}
\label{fig:vis_oscillation}
\end{center}
\end{figure}

Figure~\ref{fig:vis_oscillation} pictorially shows this process using the angular wavenumber versus frequency space $I(\u,f)$. In the cartoon the corrugated shading represents real (or imaginary) component of the emission from a single flat-spectrum source away from the beam center. An offset $\delta$-function in image space (one point source) becomes an angular corrugation after the angular FT, with constant emission strength as a function of frequency. The visibilities measured the sky emission at the baseline separations indicated by the diagonal black lines. The distance between two antennas measured in wavelengths increases with frequency ($\u_{ij} = \Delta \r_{ij}\, f/ c$), naturally leading to oscillating visibility measurements with frequency (Equation~\ref{eq:vimage2}). As baselines become longer the speed of oscillation from the foreground source increases. This is the fundamental source of the foreground wedge, as studied in depth in \citet{Vedantham:2011wu,Morales:2012uk,Trott:2012fb,Parsons2012,Hazelton:2013fu}. However, as we will explore in \S\ref{sec:foreground}, the foreground wedge does appear differently in the two families of PS estimators.

While the foregrounds are spectrally smooth, the cosmological signal has frequency structure due to redshifted line-of-sight fluctuations in the 21~cm brightness temperature. The true power spectrum of the sky is equal to taking the true sky $I(\u,f)$, mapping $\u \rightarrow \kperp$ with the angular diameter distance and mapping $f$ to line-of-sight distance and Fourier transforming to obtain $I(\kperp,\kpar)$ and squaring to obtain $P^T(\k)$ \citep{Morales2004}. However, no instrument measures the true sky. The analysis challenge is how to construct an estimate of the {\it true} PS from the {\it measured} interferometric visibilities, and it is how this estimate is constructed that determines to which family the PS analysis belongs.


%
%
%
%

\section{Power Spectrum Estimators}
\label{sec:estimators}

There have been a number of different proposed methods for estimating the cosmological PS $\Ph$ from interferometric measurements, including the imaging PS, the variance statistic, the delay spectrum, and m\nobreakdash-mode analysis \citep{Zaldarriaga2004,Morales2004,Mellema2006,Stuart2008,Iliev2008,Chang2008,Pen2008a,Vedantham:2011wu,Parsons2012,Masui2013,Switzer2013,Patil2014,Shaw2014,Dillon2014a,Paul2014,Paul2016,Trott2016,Liu2016,Gehlot2017}. While the language used to describe the estimators is highly variable, the mathematics fall into two families depending on whether they measure the PS of the measured visibilities (`delay spectrum' approaches) or the PS of the reconstructed sky (`imaging' PS approaches). In this section we carefully review the proposed measures, and sort the proposals into two families of estimators.

While complete analysis pipelines that transform raw visibilities into PS estimates are often broken into several distinct software packages, here we are considering the complete pipelines and classifying them by the characteristics of the final estimator.


%


\subsection{Imaging or Reconstructed Sky PS}
The three dimensional PS of reconstructed 21~cm EoR images was first introduced in \cite{Morales2004}. Conceptually this is a very simple estimator---it is just the 3D Fourier transform and square of the reconstructed sky estimate $\Ih(\th,f)$ mapped to cosmological coordinates. The resulting power spectrum estimator is $\Pkp$, where $\kperp$ and $\kpar$ are the angular and line-of-sight wavenumbers respectively.

\begin{figure}
\begin{center}
\includegraphics[width=\columnwidth]{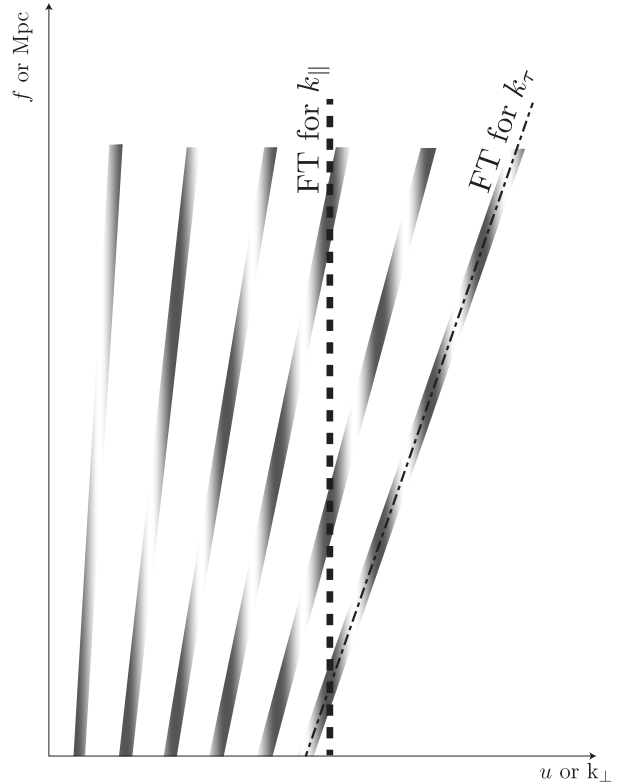}
\caption{This cartoon shows the oscillations of the different baselines in Figure~\ref{fig:vis_oscillation}, showing the faster oscillation for longer baselines. Reconstructed PS estimators $\Pkp$ take the frequency FT along a fixed angular scale, as shown by the vertical thick dashed line, whereas measured sky PS estimators $\Pkt$ take the frequency FT along the direction of the baselines as shown by the thin dash-dot line. This figure is describe in detail in \S\ref{sec:estimators}.
}
\label{fig:ps}
\end{center}
\end{figure}


The subtlety is in how the sky estimate is made from interferometric visibility measurements. Because $\kpar$ is calculated by taking the Fourier transform of $\Ih(\kperp,f)$ perpendicular to $\kperp$, the upper and lower frequencies incorporate information from baselines of different physical length as indicated by the vertical thick dashed line in Figure~\ref{fig:ps}. All of the baselines agree on the amplitude and phase of the foreground sources at that scale (dark color where the dashed line crosses), so the majority of the power after the line-of-sight FT is at $\kpar = 0$. If the estimate of the sky $\Ih$ was equal to the true sky, the imaging PS estimator $\Pkp$ would be free of foreground contamination except at the lowest $\kpar$ scales. But there are necessarily errors in the estimate of the sky, and these errors leak power from the smooth spectrum foregrounds into the cosmological volume. This leakage appears in a distinctive `wedge' pattern with a `window', that we will explore in \S\ref{sec:foreground}.

A deep literature on how to estimate $\Ih$ and $\Pkp$ from interferometric visibilities has been developed, and it has spurred the application of optimal map making and advanced data inversion techniques to interferometric data (e.g.\ \citealt{Myers2003,Bhatnagar:2008gn,Morales2009,Sullivan2012,Dillon2014a,Shaw2014,Zheng2017}). The reconstructed PS have two defining features:
\begin{itemize}
  \item The line-of-sight coordinate $\kpar$ is orthogonal to the $\kperp$ coordinate (by construction).
  \item Because of this choice of coordinates, data from multiple baselines are linearly combined in forming the values of the $\Pkp$ estimator. The Fourier transform of frequency is at a fixed angular scale, and due to the change in array size with frequency (in wavelengths) this involves incorporating information from multiple baselines in a frequency dependent way. This strongly affects how foregrounds appear in the resulting PS and the sensitivity to calibration errors, as we'll discuss in \S\ref{sec:foreground} \& \S\ref{sec:calibration}.
\end{itemize}
It is interesting to note that many of the advanced PS estimation techniques never create an image. But by creating a PS estimate in the $\kperp,\kpar$ coordinates they are implicitly working with a reconstructed sky and are combining data from baselines of different length in a frequency dependent way during the estimation process.


\subsection{Delay spectrum or Measured Sky PS}

The delay spectrum was introduced by \cite{Parsons2012} and has been used with PAPER data to provide PS limits. In the delay PS a visibility---either a single baseline or an average of redundant baselines---is Fourier Transformed along frequency and squared to form the PS estimator. Fundamentally this is a PS of the {\it measured} sky, and no reconstruction of the true sky is attempted. 

From Figure~\ref{fig:ps}, if $\kpar$ is the line-of-sight wavenumber we can see that for short baselines the visibility as a function of frequency is sampling sky information {\it nearly} parallel to the line-of-sight, but not quite. The estimator is really along $\kt$ (diagonal dash-dot line), where $\tau$ is the time delay of the electric field propagation between antennas $\tau_{ij} = \Delta \r_{ij}\cdot\th/c$, and it is this time delay that gives the delay spectrum approach its name. 

Measurements of multiple baseline lengths can be used to fill out the delay spectrum estimator $\Pkt$, and baselines of different orientation or from different sidereal times can be added incoherently to improve the signal-to-noise of the estimator. Significant additional work has improved the sensitivity of the technique by optimal time integration fringe-rate filtering \citep{Parsons2016}, leading to a very deep literature and mature estimator. But for our purposes the measured sky PS estimators have two key features:
\begin{itemize}
  \item Because the Fourier Transform from $f$ to $\kt$ follows the baseline migration (diagonal lines in Figure~\ref{fig:ps}), the same baselines contribute at all frequencies. Redundant baselines can be added together to create a single low-noise measurement before the Fourier Transform and squaring operations \citep{Parsons2012}, and advanced versions can combine baselines that are not identical \citep{Zhang2017}. But crucially, the baselines contributing to the Fourier Transform do not change with frequency.
  \item The line-of-sight coordinate $\kt$ is nearly, but not exactly, orthogonal to the transverse coordinate $\kperp$. Because $\kt \approx \kpar$ this has a negligible effect on the cosmological PS, and if needed the theory could be rephrased in terms of $\Pkt$ \citep{Parsons2012}. However, choosing $\kt$ as the basis of the estimator has a significant impact on how the foreground emission contaminates the PS estimate as explored at length in \S\ref{sec:foreground}.
\end{itemize}



\subsection{Classifying Analysis Efforts}

\begin{table*}[t]
\centering
\renewcommand{\arraystretch}{1.5}
\begin{tabular}{|c|c|}
\hline
    \bf{Measured PS} $\Pkt$ & \bf{Reconstructed PS} $\Pkp$ \\
    \hline \hline \renewcommand{\arraystretch}{1.0}
    PAPER \& HERA Delay PS  & GMRT  \\ 
    (\citealt{Parsons2012,Ali2015}; & \citep{Paciga2011} \\
    \citealt{Jacobs2015,Kohn2018}) & \\
    \hline
    MWA Gridded Delay PS  & GBT auto \& cross correlation \\ 
    \citep{Vedantham:2011wu,Paul2016} &  \citep{Masui2013,Switzer2013} \\ 
    \hline
    LOFAR Gridded Delay PS \citep{Gehlot2017} & LOFAR variance \citep{Patil2014} \\ 
    \hline
    & LOFAR PS \citep{Patil2017} \\ 
    \hline
    & MWA FHD + EmpCov \\
    & \citep{Dillon2015,Ewall-Wice2016} \\ 
    \hline
    & MWA FHD + \eppsilon\\
    & (\citealt{Jacobs2016,Beardsley2017})\\ 
    \hline
    & MWA RTS + CHIPS \\
    & (\citealt{Trott2016,Beardsley2017}) \\ 
    \hline
    & CHIME m-mode \citep{Shaw2014,Shaw2015} \\
\hline
\end{tabular}
\caption{Major 21~cm PS analysis efforts.}
\label{table:classification}
\end{table*}

Table~\ref{table:classification} classifies the 21~cm PS analysis efforts that have been developed into full analysis pipelines. The first column lists the measured sky PS ($\Pkt$), and the second column lists the reconstructed sky PS ($\Pkp$). Every one of these analysis efforts is backed by many more papers than we have space to list here (see references therein). Other analyses such as \citet{Peterson2009} and \citet{Liu2016} can also be sorted into $\Pkt$ or $\Pkp$ families, but have not yet been developed in to full data-ready software pipelines. 

The process of classifying all of the analyses highlights two particularly interesting variants:  the m\nobreakdash-mode analysis by \citet{Shaw2014,Shaw2015} and the physical gridding proposal by \cite{Vedantham:2011wu} and \cite{Paul2014,Paul2016}.

The m\nobreakdash-mode analysis was proposed by \citet{Shaw2014} and further developed in \citet{Shaw2015}, and is the basis for the CHIME  Dark Energy  measurement. On first glance the m\nobreakdash-mode formalism looks very different from most PS analyses, and in many ways it is:  no image is ever formed, it is optimized for a drift telescope that sees all right ascensions, and it must work deep in the foreground wedge to achieve CHIME's dark energy science. In the m\nobreakdash-mode analysis the antenna beams are decomposed into spherical harmonics, and are used to map the measured visibilities directly into the $l,m$ spherical harmonic basis. Further, in CHIME the close packing of the antennas along the feed in the N-S ($l$) direction and the rapid sampling in time ($m$) leads to almost perfect sampling. Because of the drift scanning strategy that covers all right ascension, these measurements are never mapped to a `uv-plane'. But in the chosen harmonics the measurement coverage is nearly perfect up to the angular resolution of the instrument---perfect `uv' coverage. 

While the m\nobreakdash-mode analysis never creates an image, it does estimate the $\hat{a}_{lm}$ coefficients of the sky, and is properly classified as a reconstructed sky PS. The estimator is explicitly in the $\Pkp$ frame, and if desired the $\hat{a}_{lm}$ coefficients could be transformed into a reconstructed image of the sky. Further, because the instrument is chromatic there is a frequency-dependent mapping of baselines to $\hat{a}_{lm}(\kpar)$ estimators---baselines of different physical length are linearly combined to form one estimator. These are the two hallmarks of a `reconstructed sky' PS analysis: using the $\Pkp$ basis and combining data from baselines of different physical size in a frequency dependent manner. m\nobreakdash-mode analysis is in the same family as more traditional imaging PS, albeit a very advanced implementation specifically tuned for a unique style of instrument.

Another variant first proposed by \citet{Vedantham:2011wu} and further developed by \cite{Paul2014,Paul2016} and \cite{Gehlot2017}, is to grid visibilities by their physical separation ($\r$ instead of $\u$). This can be visualized as a skewed version of Figure~\ref{fig:vis_oscillation}, where the black baselines now run vertically but the fringes from a flat spectrum source run diagonally (same crossing angle, fringes run from top left to bottom right). One can efficiently image this kind of analysis using the Chirp-Z Transform in place of the Fourier Transform, but the proposed PS estimator is to take the Fourier transform along the black lines of set physical separation. While gridding is normally associated with imaging and reconstructing a sky, in this case the PS estimator is implicitly in $\Pkt$, and because the FT aligns with the physical baseline separation, data from baselines of different physical length are never combined in a frequency-dependent way. So despite the gridding, this matches the characteristics of a measured sky delay spectrum analysis. 

Which family an estimator belongs to is of more than academic interest, as it directly determines how foregrounds and calibration errors affect the final PS measurement. In general, any lesson learned can be directly applied to all of the analyses in the same family. There may need to be some translation, but even technical effects such as $uv$ beam clipping (how the contribution of a baseline goes to zero as it passes away from the angular mode of interest, e.g.\ Figures~\ref{fig:ps}~\&~\ref{fig:calibration}) translate generally from FHD to CHIPS and m-mode analysis and all of the other $\Pkp$-style analyses. Similarly the `pitchfork' effect identified by \citet{Thyagarajan2015}  can be directly translated to all of the measured sky $\Pkt$ analyses. 

In the next two sections we will explore how residual foregrounds and calibration errors generically affect analyses within the two families.



\section{Foreground features}
\label{sec:foreground}

One of the significant developments in 21~cm Cosmology has been the recognition and understanding of the `Foreground Wedge' and the associated `EoR Window'. The appearance of the foreground wedge in both reconstructed and measured PS has led to the erroneous conclusion that the approaches are effectively identical. While the foreground wedge appears in the same {\it location} in both formalisms, the {\it amplitude and characteristics} of the wedge differ between them. 

\begin{figure*}[t]
\begin{center}
\includegraphics[width=\textwidth]{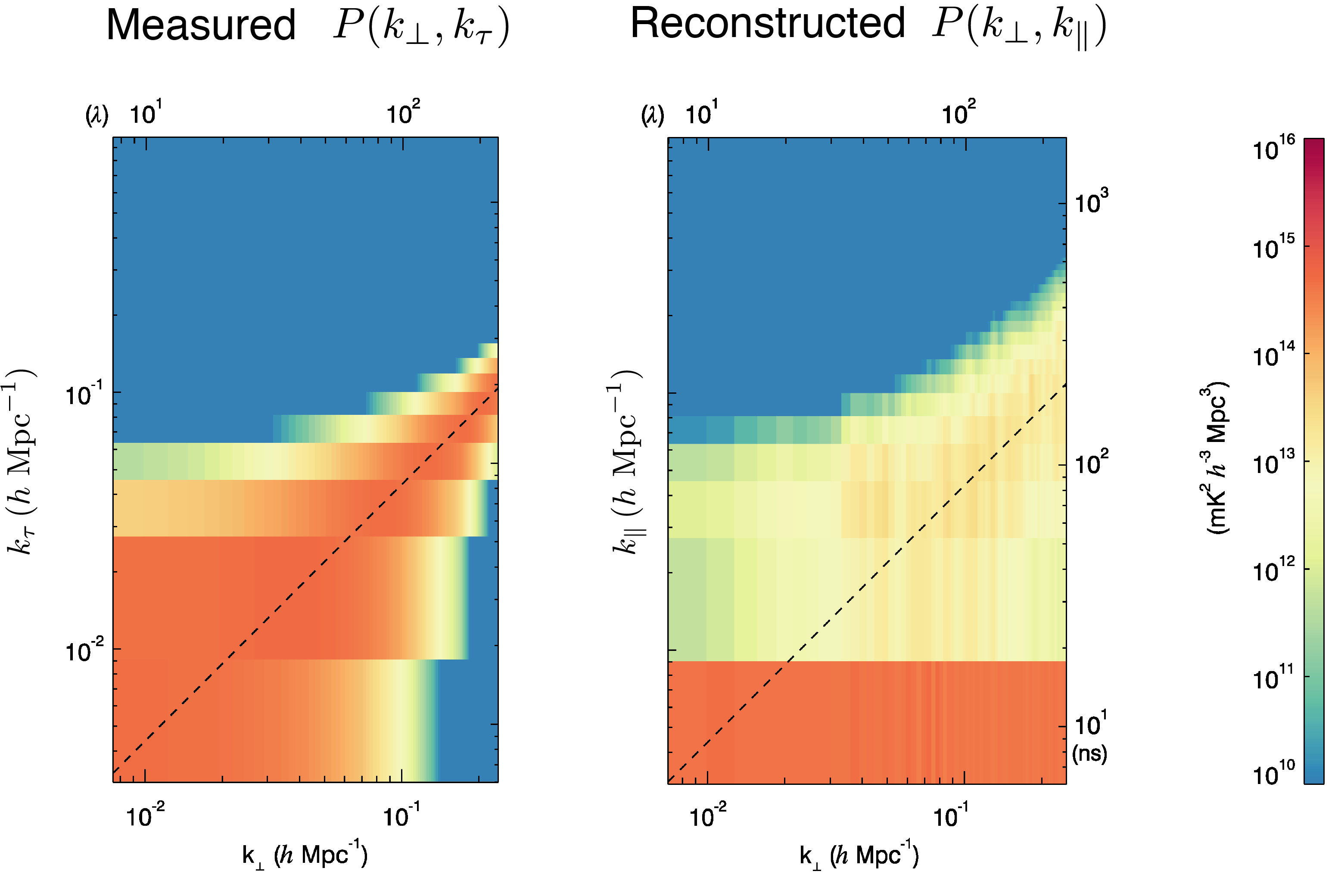}
\caption{This figure shows the measured (left) and reconstructed (right) PS for the same simulated data. Visibilities for a single flat-spectrum foreground source near the edge of the field-of-view were analytically generated for an instrument with uniform baseline density. 
In both PS we have taken a 2D slice through the 3D PS volume with $\kperp$ aligned with the source. In the lefthand measured PS all of the power from the contaminating source appears at the associated band in the PS (the dashed line shows the theoretical contamination location of the source). In the righthand reconstructed PS most of the power is reconstructed at $\kpar = 0$, with a smaller amount of power appearing in the same diagonal band due to imperfect sky reconstruction. While both PS have foreground contributions at the same location in the wedge, the amplitude of that contamination and the power at $\kpar \approx 0 $ is quite different as described in the text. If we were to add many more sources at different locations the wedge would fill in as seen in data.}
\label{fig:delayVSimaging}
\end{center}
\end{figure*}

To illustrate these effects we have analytically calculated the visibilities (Equation~\ref{eq:vimage2}) for a single flat spectrum source towards the edge of the field of view for an idealized array 
with uniform $uv$-density.\footnote{Uniform $uv$-density is not realizable with a physical antenna layout, but is useful for illustration purposes as the baseline density does not depend on $\kperp$.} We then use these analytic visibilities with full PS estimation pipelines to produce both measured and reconstructed sky PS shown in Figure~\ref{fig:delayVSimaging}.\footnote{At the dynamic range shown all measured PS estimators will give the left hand plot and all reconstructed PS estimators will give the right hand plot. At higher dynamic ranges implementation details will become important. For these figures the MWA beam was used to simulate the visibilities, a custom python implementation of the delay PS was used for the lefthand plot, and the FHD-\eppsilon\ reconstruction package was used to create the right hand plot (detailed in Barry et al., in prep.). Both incorporate a Blackman-Harris spectral window function.} 


Referring back at Equation~\ref{eq:vimage2}, we can see that a single source will cause frequency ripple in a visibility proportional to $e^{-i2\pi [\Delta \r_{ij}\, f/c] \cdot \th} = e^{-i2\pi \u_{ij}(f) \cdot \th_S}$ where $\u$ is proportional to the angular cosmological distance $\mathbf{k}_\bot$. For the measured PS we simply Fourier transform each visibility along frequency to get a $\delta$-function in delay (convolved with a spectral window function associated with the bandpass and frequency-dependent beam, \citealt{Parsons2012}). For our  array with many baselines and one contaminating source, the resulting measured sky PS $\Pkt$ is shown in the left panel of Figure~\ref{fig:delayVSimaging}. The power from the contaminating source appears at a constant diagonal line at $\tau \propto \mathbf{k}_\bot \cdot \th_S$. 

The primary feature of the foreground contamination in the measured sky PS is that {\bf all} of the power from a contaminating source appears at the associated $\kt$. The delay transform is a form of one dimensional `imaging', with every source having a specific delay. With many sources at different locations the wedge will fill in via superposition, and because it is the apparent brightness of the sources that matters the amplitude of the foreground wedge will fall in $\kt$ proportional to the antenna beam with no meaningful contamination beyond the earth's horizon (\citealt{Thyagarajan2015} carefully explores how diffuse structure appears near the horizon in measured sky PS). As the foreground sources are approximately four orders-of-magnitude brighter than the EoR signal, the contamination of the foreground in the wedge is much brighter than the expected signal. In data this bright foreground emission with a clear drop in emission at the horizon is most clearly seen in \cite{Pober:2013cp} and confirmed in simulations and observations by \cite{Thyagarajan2015,Thyagarajan2015a}.

For reconstructed sky PS, the frequency Fourier transform is taken at a fixed angular scale ($\u$) instead of along individual visibilities. Often this is accomplished by gridding to a $uvf$-cube prior to the Fourier transform, but the same effects can be incorporated with frequency-dependent visibility mappings such as in the m\nobreakdash-mode and CHIPS analyses. This transform at a fixed angular scale is shown visually by the thick dashed vertical line in  Figure~\ref{fig:ps}. Along that line the different baselines agree on the amplitude and phase of a contaminating source, so the dominant term in the Fourier transform is the flat spectrum $\kpar = 0$ DC mode. The righthand panel of Figure~\ref{fig:delayVSimaging} shows the reconstructed sky PS for the same analytic visibilities. There is contamination along the same diagonal band as seen in the lefthand measured PS, but at a significantly reduced amplitude. 

The amplitude of the foreground wedge in a reconstructed PS is related to the accuracy of the sky reconstruction---with perfect reconstruction all of the foreground emission is correctly mapped to $\kpar \approx 0$ and the foreground wedge disappears. In practice the accuracy of the sky reconstruction depends on both the analysis and the instrument characteristics. For smooth spectrum sources the instrumental effects indicated in Figure~\ref{fig:vis_oscillation} can be inverted to create a catalog or model, either through traditional deconvolution (e.g.\ \citealt{Carroll2016,Hurley-Walker2017}) or more formal matrix inversion techniques (e.g.\ \citealt{Shaw2014}). Foreground emission captured in these models can be fully removed (modulo calibration \S\ref{sec:calibration} and instrument model errors), with no associated emission in the foreground wedge. Residual foreground emission not captured in these models has its power split between the $\kpar \approx 0$ modes and the wedge as shown in the righthand panel of Figure~\ref{fig:delayVSimaging}. The fraction of the  residual emission in the diagonal band is determined by the point spread function (PSF) of the instrument---the better the PSF the more accurate the (dirty) sky reconstruction and correspondingly less emission in the foreground wedge. As the $uv$ coverage and calibration precision of an array increases, both the PSF and the depth of foreground source catalogs improve and the magnitude of the emission in the foreground wedge decreases.

While both measured and reconstructed sky PS see a foreground wedge, the wedge has qualitatively different properties:
\begin{itemize}
  \item For the measured sky PS the foreground naturally reconstructs to $\mathbf{k}_\bot \cdot \th_S$. $\kt = 0$ simply refers to sources at phase center (zero delay) and is not preferred over other locations in the wedge. The amplitude of the foreground emission in the wedge is much brighter in a measured PS, because it contains the full amplitude of the foregrounds. This is clearly seen in \cite{Pober:2013cp} and \cite{Thyagarajan2015a}.
  \item For the reconstructed sky PS, most of the foreground power appears at $\kpar \approx 0$, with leakage due to imperfect foreground isolation forming a line at the associated $\mathbf{k}_\bot \cdot \th_S$. The wedge will fill in with the superposition of many sources but the amplitude of the wedge is strongly associated with the instrument PSF and the quality of the foreground removal. This is seen in PS measurements in \cite{Jacobs2016} and \cite{Beardsley2016}.
\end{itemize}

Measured and reconstructed PS estimators are asking different statistical questions. The measured PS estimator $\Pkt$ is the power spectrum of the calibrated visibilities, and does not explicitly project sources to a specific location on the sky. A particular $\kt$ in the wedge can be generated by either a flat spectrum source at a particular position {\it or} by sources with spectral variations at other locations---the estimator does not distinguish. In contrast, $\Pkp$ is the PS of the reconstructed sky. The reconstruction actively tries to remove the effects of the chromatic instrument and place the emission at its true location. The errors in this reconstruction due to unsubtracted sources coupling with the chromatic PSF lead to the emission in the wedge.  In both estimators the foreground wedge is generated by the baseline movement of the instrument (Figure~\ref{fig:vis_oscillation}), so the wedge appears in the same location. But because the estimators are asking different questions the foreground characteristics and the amplitude of the wedge are different.


The differences in how the foregrounds appear in the two classes of estimators impact the kinds of science that can be pursued. But to fully understand these effects we must also understand how the analyses respond to calibration errors. 


\section{Calibration}
\label{sec:calibration}
%
The importance of calibration has long been recognized in 21~cm cosmology, and is a very active area of current research (e.g.\ \citealt{Barry2016,Trott2016,Patil2016,Ewall-Wice2017}). The purpose of calibration is to measure the as-built instrument performance---$A_i(\th,f)$ in Equation~\ref{eq:vimage}---and include it in the analysis. A full understanding of the causes and effects of calibration errors is well beyond the scope of this paper, however, it is instructive to qualitatively understand how calibration errors affect our two families of PS estimators.


 Because calibration is tightly associated with array layout, we find the clearest general argument is analytic. Assuming the broad calibration worked well, we will focus only on small differences between the true antenna performance and the antenna calibration used in the analyses---the calibration errors. In the limit of small calibration errors it is conceptually helpful to factor the error in the antenna model into three terms:
\beqa
\Delta A_i(\th,f) = \overbrace{\Delta a_i e^{i\Delta\phi_i}}^{\rm f\ indep.} \times \overbrace{\Delta a_i(f)e^{i\Delta\phi_i(f)}}^{\rm f\ dep.} \times \overbrace{\Delta G_i(\theta,f)}^{\rm beam\ shape} \nonumber \\
\label{eq:calError}
\eeqa 
These terms represent the overall frequency-independent error in the sensitivity ($\Delta a$) and timing ($\Delta \phi$) of an antenna; the frequency-dependent sensitivity and timing errors; and the beam shape error. In the following subsections we'll explore how these different kinds of calibration errors affect the reconstructed and measured sky PS. The major result of this analysis is that reconstructed PS are much more sensitive to frequency-independent calibration errors than measured PS.

\subsection{Frequency-independent calibration errors}
\label{sec:findep}

Our first case is a small frequency-independent error in either the sensitivity or phase of one antenna ($\Delta a$ or $\Delta \phi$).
For the delay spectrum of a single baseline, frequency-independent calibration errors have almost no effect. The amplitude makes a minute error in the strength of the PS. The phase error corresponds to a small shift in $\tau$. If many baselines are added together incoherently the phase error will shift and blur the delay PS very slightly, but will have no material effect on the location and amplitude of the foreground emission and the EoR window should remain clean.

\begin{figure}[t]
\begin{center}
\includegraphics[width= \columnwidth]{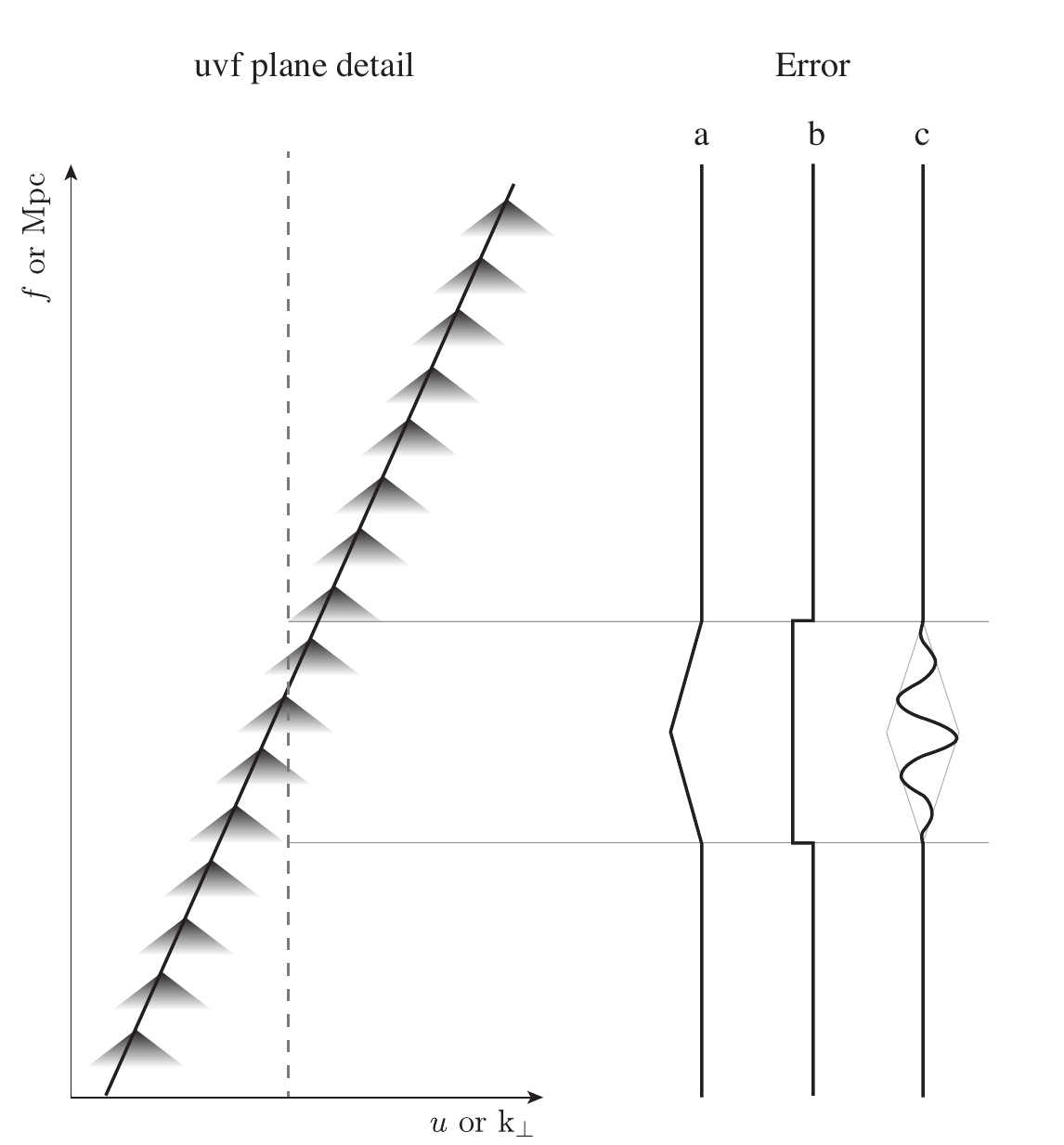}
\caption{The contribution of a single miscalibrated baseline to the reconstructed $uvf$ cube. In the left panel a single miscalibrated baseline (diagonal line) contributes to the estimate at a particular angular scale (vertical dashed line) over a range of frequencies. At every frequency the baseline contributes to a range of angular scales as indicated by the small triangles (the gridding kernel in an imaging analysis or the covariance scale in other reconstructed sky estimators). The error contribution of the miscalibrated visibility to the angular scale of interest is shown to the right in three scenarios:  a) frequency-independent calibration error in the high baseline number limit, b) frequency-independent calibration error in the single baseline limit, and c) frequency-dependent calibration error in the high baseline number limit. }
\label{fig:calibration}
\end{center}
\end{figure}

In contrast, for a reconstructed sky PS the same calibration error can move power from the $\kpar = 0$ line into the foreground wedge and potentially beyond. 
%
We can show this pictorially in Figure~\ref{fig:calibration} for simple interferometric imaging, though the same mathematical effect occurs in all reconstructed PS, even those in which no image is explicitly formed. In simple interferometric imaging a visibility is first gridded to the $uvf$-cube ($\Ih(\u,f)$) using a gridding kernel with a small width in $\Delta u \Delta v$, as shown in the lefthand side of Figure~\ref{fig:calibration}.
Mathematically the estimated apparent sky in $uvf$ coordinates is
\beq 
\Ih(\u,f) =  \frac{\sum_{ij}v_{ij}(f)B^*(\u-\Delta\r_{ij}f/c,f)}{\sum_{ij}B^*(\u-\Delta\r_{ij}f/c,f)},
\label{eq:calConv}
\eeq
where $B^*$ is the gridding kernel. The term $\Delta \r_{ij}f/c$ captures the frequency-dependent movement of the baseline. Because of the finite width of the gridding kernel, each baseline contributes to a range of frequencies with a weight associated with the $uv$-space beam shape. The range of frequencies depends on the size of the gridding kernel (antenna size, inverse of FoV), and the baseline length since longer baselines cross more quickly. We see that along any given angular scale $\u$ the $uvf$-cube $\Ih(\u,f)$ is a weighted average of the visibilities. 

The question is how are $\Ih(\u,f)$ and $\Pkp$ affected by a single miscalibrated visibility measurement. Because $\Ih(\u,f)$ is a weighted average, the effect depends on how many visibilities are contributing. 

In the limit of many visibilities, a frequency-independent error takes on the shape of the $uv$-beam as it crosses the angular scale $\u$ of interest, as shown in a) on the righthand side of Figure~\ref{fig:calibration}. This produces an enveloped error. The shape of the envelope depends on the gridding kernel, and how close the baseline comes to directly crossing the angular scale $\u$ (impact parameter), but for a single miscalibrated baseline that does directly cross and an angularly symmetric frequency-independent beam, the error in the reconstruction at the angular scale $\u$ is 
\beq 
\Delta\Ih(\u,f) =  \Delta v_{ij}B^*(\Delta f\,\Delta\r_{ij}/c),
\eeq
where $\Delta f$ is the difference from the frequency at the center of the baseline crossing.
The frequency shape of the error along the angular scale of interest is just the shape of the gridding kernel times a scaling factor $\r_{ij}/c$ related to how quickly the baseline moves with frequency. In most analyses the gridding kernel is the angular FT of the antenna beam shape, so the contamination is a scaled version of the antenna beam in $uv$ coordinates. In a nice bit of symmetry, when we perform the frequency FT $f \rightarrow \kpar$ to calculate $\Pkp$, this translates the $uv$ beam back to scaled angular coordinates. The calibration contamination in the PS is enveloped by a scaled version of the angular antenna beam, centered at $\kpar = 0$. So in the limit of many baselines, frequency-independent calibration errors move power from $\kpar \approx 0$ into the wedge for a reconstructed PS. (A full mathematical description can be obtained by substituting in the visibility error $\Delta v$ into the mathematics of \citealt{Liu2014a}, and can be qualitatively understood with a similar substitution into \citealt{Hazelton:2013fu} and the animated version their Figure~2 available online.)

There is however, the other limit of very few baselines. In this limit the weighting provided by the beam in the numerator and denominator of Equation~\ref{eq:calConv} divide out, and we get a sharp change in the reconstructed $\Ih(\u,f)$ as the miscalibrated visibility crosses, as shown in case b) of Figure~\ref{fig:calibration}. This sharp change leads to calibration errors contributing to all $\kpar$, including those in the window. 

These effects can be seen in data. For the MWA, the baseline density is very high at small $\kperp$ and becomes sparse at large $\kperp$. In the PS published in \cite{Jacobs2016} and \cite{Beardsley2016}, the contamination at low $\kperp$ is mostly confined to the wedge because the baseline density is high, whereas it leaks into the window at high $\kperp$ where the baseline density becomes low and the opposite limit is reached. 

Sparse baselines usually correspond to low sensitivity, so only areas of high baseline density materially contribute to most PS measurements. So the frequency-independent calibration errors are confined to the wedge for most reconstructed PS. However, the baseline density is not infinite, so some of the contamination can bleed into the window.

The frequency-dependent contribution of a measurement to a single angular scale---the hallmark of a reconstructed sky PS estimator---is what causes frequency-independent calibration errors to move power from $\kpar\approx0$ into the wedge and potentially into the window. This is fundamental to the approach, as it was exactly this frequency-dependent baseline contribution that enabled the power to be reconstructed to $\kpar\approx0$ in the first place. This kind of calibration error, where baselines of different physical length must agree very precisely on the amplitude and phase, is a very active area of current research. 

In conclusion, measured sky PS estimators $\Pkt$ are much less sensitive to  frequency-independent calibration errors than reconstructed sky PS estimators.

\subsection{Frequency-dependent calibration errors}
\label{sec:fdep}

There are two other classes of calibration errors from Equation~\ref{eq:calError}. The first of these is frequency-dependent calibration errors. Effects like a cable reflection can source sinusoidal ripples in the antenna sensitivity (and phase) that are nearly impossible to calibrate at the necessary precision \citep{Beardsley2016,Barry2016}. These frequency-dependent calibration errors can be described with a set of $\delta$-functions in $\tau$ at the corresponding time delays. For the delay spectrum these errors act like a large additive delay $\Delta \tau$ and can directly move power from the foreground wedge into the window. 

For a reconstructed sky analysis in the large baseline limit, the impact of a sinusoidal calibration error is shown in c) on the righthands side of Figure~\ref{fig:calibration}. The sinusoidal calibration error appears, enveloped by the gridding kernel. This will throw power into the EoR window, as in the measured PS case, but it will not be as sharp a contribution due to the modulation of the sinusoid by the gridding kernel. The effect of frequency-dependent calibration errors on reconstructed PS was simulated and explored in depth in \citet{Barry2016}.


As both classes of PS estimators are extremely sensitive to frequency-dependent calibration errors, eliminating  spectral features is driving the instrumental requirements of HERA, MWA~III, and SKA-Low \citep{DeBoer2016}.

\subsection{Beam shape errors}
\label{sec:beamshape}

The last calibration error from Equestion~\ref{eq:calError} is the antenna beam shape. In particular, due to manufacturing and electronic differences one antenna may have a different angular response. Measuring the antenna beam using celestial sources (e.g.\ \citealt{Newburgh2014,Berger2016}), satellites (\citealt{Neben2015,Neben2016,Neben2016a}, Line et al.\ in press), and drones \citep{Jacobs2017} is an active area of research.

Using modern techniques (e.g.\ \citealt{Bhatnagar:2008gn,Morales2009,Myers2003}) sky reconstruction analyses can correct for the direction-dependent polarized response of individual antennas. Effectively, the analysis can project visibilities to the sky and apply direction-dependent corrections during image reconstruction. The PS of the reconstructed image is then free of these errors to the precision of the direction-dependent calibration.

In contrast, because measured sky PS have no sky reconstruction step it is more difficult to apply a direction-dependent gain correction. \citet{Parsons2016} shows how limited direction-dependent information can be propagated to the PS estimator, but there is no way to completely correct for angular differences in the polarized antenna beams. The measured sky PS inherently assume the angular gain of all antennas are similar and that the polarization at the relevant spectral scales is small (e.g.\ \citealt{Moore2013a,Kohn2016,Moore2017}). 

In both classes of analyses purely angular differences in the antenna beam shape will be confined to the wedge. This can be seen by substituting perturbed beams into \cite{Liu2014a}, or by noting that beam errors will make certain areas of the sky appear artificially bright or dim but not affect the spectral structure. In reconstructed analyses the contamination of the wedge can be reduced by using knowledge of the individual antenna beams to improve the reconstruction. There is no fully analogous technique for measured PS, but the foreground wedge is already extremely bright for these measurements.


\subsection{Calibration effects reviewed}

The two classes of PS estimators have different responses to the three kinds of calibration errors described in Equation~\ref{eq:calError}.
\begin{itemize}
  \item Both measured and reconstructed PS are extremely sensitive to frequency-dependent calibration errors (\S\ref{sec:fdep}). These errors can move foreground power into the window. This sensitivity to spectral ripple is driving the intrinsically smooth bandpass requirements for antennas in next generation 21~cm instruments.
  \item Frequency-independent calibration errors (\S\ref{sec:findep}) have negligible impact on measured PS, but in reconstructed PS can move power from $\kpar \approx 0$ into the wedge and window. Amplitude and phase calibration between baselines of different physical length must be very precise to enable working within the wedge. 
  \item Direction-dependent calibration can be incorporated in reconstructed sky PS, but there is no natural way of incorporating direction-dependent effects in measured sky PS. 
\end{itemize}

\section{Discussion}
\label{sec:ProsCons}

By realizing that all of the proposed 21~cm analyses can be categorized into two broad classes of PS estimator (\S\ref{sec:estimators}, Table~\ref{table:classification}), we can start to ask more generic questions about the different approaches. In this paper we've identified two major features: the brightness of the wedge (\S\ref{sec:foreground}) and the sensitivity to frequency-independent calibration errors (\S\ref{sec:findep}). 

The location of the foreground wedge is determined by the chromatic nature of an interferometer (Figure~\ref{fig:vis_oscillation}), and is the same for both measured and reconstructed PS estimators. However, the wedge is much brighter for measured PS estimators. Because measured PS analyses make no attempt to reconstruct the true location of the emission, all of the foreground power in a visibility is mapped to the corresponding location in the wedge (Figure~\ref{fig:delayVSimaging}a). In contrast, for reconstructed PS estimators most of the smooth spectrum emission is mapped to $\kpar\approx0$ (Figures~\ref{fig:ps}~\&~\ref{fig:delayVSimaging}b). The more precise the reconstruction, the less power appears within the wedge. Reconstruction errors due to residual sources, instrumental PSF, and calibration errors determine the brightness of the foreground wedge in a reconstructed PS analysis.

A particular advantage of measured PS estimators is their immunity to frequency-independent calibration errors (\S\ref{sec:findep}). Because the frequency Fourier Transform is taken along $\kt$ (Figure~\ref{fig:ps}), frequency-independent calibration errors have almost no effect on the resulting PS. In contrast, because different baselines contribute at different frequencies for the FT to $\kpar$, reconstructed PS are quite sensitive to frequency-independent calibration errors (Figures~\ref{fig:ps}~\&~\ref{fig:calibration}, \S\ref{sec:findep}).

For analyses that limit themselves to the window, there is no clear advantage to  either the measured or reconstructed PS approaches---contamination in the window is primarily associated with frequency-dependent calibration errors. Instrument specific differences such as the smoothness of the bandpass, antenna layout, and calibration approach quickly determine the precision that can be reached in the window. 

Things become more interesting if one wishes to work within the wedge. Both LOFAR and CHIME are predicated on being able to work deep within the wedge, and there is a large boost in both sensitivity and science reach that can be obtained if the cosmological modes within the wedge can be used (e.g.\ \citealt{Pober:2013cp}). 

In practice working within the wedge requires a reconstructed PS analysis. PS of a reconstructed sky have the advantage that the foreground contamination within the wedge is attenuated (Figure~\ref{fig:delayVSimaging}b). In contrast, the full foreground emission appears in the measured PS wedge (Figure~\ref{fig:delayVSimaging}a). Working within the wedge with a measured PS would require much more precise foreground models because there is no natural suppression from the instrumental PSF. Worse, determining the foreground model to subtract would require an imaging analysis. Within the delay spectrum approach there is no way to identify whether the flux in a $k$-space bin is foreground or cosmology---both smooth spectrum foregrounds and the spectrally varying cosmological signal map to the same bins. Measuring the foregrounds to subtract in order to work within the wedge would require an imaging analysis stage, and it would require imaging to a significantly higher precision than is needed for a reconstructed PS analysis. To work within the wedge with the measured PS one would have already been able to perform the analysis on lower precision sky estimates with a reconstructed PS estimator.

The measured PS estimators do enjoy a smaller software development burden. While all of the PS efforts involve building large custom software analysis pipelines, the complexity of the codes needed to accurately reconstruct the sky is significantly higher (e.g.\ \citealt{Sullivan2012,Trott2016,Jacobs2016}, Hazelton et al.\ in prep, Pindor et al.\ in prep). Because of the unprecedented precision required for 21~cm imaging, extraordinarily small errors can lead to unexpected deleterious results, such as those described in \citet{Trott2016,Beardsley2016,Patil2017,Barry2016}. Reconstructing the sky at the precision needed for 21~cm cosmology is incredibly difficult.

There are nascent efforts to develop hybrid analyses such as \citet{Kerrigan2018}. It may be possible to filter foregrounds in one style of analysis and construct a PS in the other, or transfer calibrations. These methods are in their infancy but crossing the divide between the measured and reconstructed PS analyses may capitalize on the unique advantages of both.

\section{Conclusion}

The GBT, PAPER, MWA, \& LOFAR teams are all actively reducing hundreds of hours of EoR data, and the progress of all the associated analyses is progressing rapidly. CHIME is just starting to collect its first science data, major hardware upgrades to the MWA and LOFAR are underway, HERA construction has begun, and SKA-low is on the horizon.

Looking to the future of 21~cm cosmology analysis, it is the belief of many in the community that we will need to incorporate lessons from all of the different approaches, and that both measured and reconstructed PS approaches have key roles to play. HERA, MWA-II, and CHIME have all been designed to have both redundant baselines (best for measured sky PS) and excellent $uv$ coverage (best for reconstructed PS), and HERA and MWA-II are explicitly planning to perform both styles of analysis.

This is an exciting time for 21~cm cosmology as we develop the foundational analysis tools and techniques. In this paper we have endeavored to conceptually organize the different proposed analyses (\S\ref{sec:estimators}), characterize how foregrounds and calibration errors appear (\S\ref{sec:foreground} \& \S\ref{sec:calibration}), and highlight the relative advantages of the different approaches (\S\ref{sec:ProsCons}).

In future work we intend to further explore the subtle differences in the shape and extent of the window between the two classes of analysis, and we hope our categorization will help best practices be more quickly shared across the many groups working on 21~cm cosmology.

\section*{Acknowledgments}

This work was directly supported by NSF grants \#1613855, \#1613040, \#1506024, and \#1636646 and NASA grant 80NSSC18K0389.
APB is supported by an NSF Astronomy and Astrophysics Postdoctoral Fellowship under \#1701440. We'd also like to thank the anonymous referee, Adrian Liu, Cathryn Trott, and Michael Eastwood for their helpful comments in improving this paper.

\bibliographystyle{aasjournal}
\bibliography{morales,library,whitebook}
\end{document}